# Evaluating the impact of next generation broadband on local business creation

Philip Chen[1], Edward J Oughton[2*], Pete Tyler[1], Mo Jia[1] and Jakub Zagdanski[3]

[1]University of Cambridge, Cambridge, UK
[2]George Mason University, Fairfax, VA
[3]Cambridge Econometrics, Cambridge, UK
*Corresponding author: Edward J. Oughton (E-mail: eoughton@gmu.edu; Address: GGS, George Mason University, 4400 University Drive, Fairfax, VA)

ABSTRACT

Basic broadband connectivity is regarded as generally having a positive macroeconomic effect. However, over the past decade there has been an emerging school of thought suggesting the impacts of upgrading to higher speed broadband have been overstated, potentially leading to the inefficient allocation of taxpayer-funded subsidies. In this analysis we model the impacts of Next Generation Access on new business creation using high-resolution panel data. After controlling for a range of factors, the results provide evidence of a small but significant negative impact of high-speed broadband on new business creation over the study period which we suggest could be due to two factors. Firstly, moving from basic to high-speed broadband provides few benefits to enable new businesses being formed. Secondly, strong price competition and market consolidation from online service providers (e.g. Amazon etc.) may be deterring new business start-ups. This analysis provides another piece of evidence to suggest that the economic impact of broadband is more nuanced than the debate has traditionally suggested. Our conjecture is that future policy decisions need to be more realistic about the potential economic impacts of broadband, including those effects that could be negative on the stock of local businesses and therefore the local tax base.

1. INTRODUCTION

Infrastructure is increasingly moving up the policy agenda as one option to help drive economic growth and productivity (Kongaut and Bohlin, 2014; Glass and Tardiff, 2019; David, 2019; Hall et al. 2016a; Hall et al. 2016b). Although various studies suggest positive macroeconomic economic effects (Koutroumpis, 2009; Czernich et al. 2011; Nadiri, Nandi and Akoz, 2018), there is a growing school of thought which argues that the economic impact of high-speed broadband is over-stated and could lead to broadband subsidies wasting billions of dollars of tax payer's money (Kenny and Kenny, 2011; Ford, 2018). While basic broadband infrastructure access is known to be a necessary ingredient to enable local economic development (Kolko, 2012; Jayakar and Park, 2013), there is debate around the level of influence high-speed broadband has on various economic metrics such as economic growth, employment and business creation.

Currently, we still lack detailed evidence on how investment in broadband infrastructure affects economic performance at the local level which has hitherto been largely overlooked (Holt and Jamison, 2009). This is particularly the case for how the spatial roll-out of new broadband technologies either enable or constrain business creation depending on availability. The broadband access agenda is now even more important considering the COVID-19 global pandemic where remote working and online activities have been a cornerstone of social distancing strategies, yet there is concern some areas are being left behind.

The analysis presented in this article seeks to understand more about how the provision of broadband affects business creation at the local level using high resolution spatial broadband data. The article defines business creation as the formation of new local business units, which have the potential to provide both additional economic value and employment to a local economy. The study focuses on the roll-out of Next Generation Access (NGA), which is a highly important topic because of the significant emphasis given to these broadband technologies in telecommunications policy over the past decade (Feijóo, Gómez-Barroso and Bohlin, 2011). NGA is indicative of premises having access to at least Fibre-To-The-Cabinet or cable DOCSIS 3.0, but also potentially higher capacity technologies such as Fibre-To-The-Premises. Using annual data from 2011 to 2015, a panel data analysis is undertaken of business establishment growth for local statistical areas.

The article is organised as follows. Section 2 summarises the relevant literature. The econometric method is described in Section 3, along with a data description in Section 4. Section 5 presents key findings and Section 6 provides concluding thoughts.

## 2. LITERATURE REVIEW

The general role of infrastructure (e.g. transportation) as an additional factor of production driving economic performance has been central to the ongoing debate since Aschauer (1989). More recently, the focus has been moving towards assessing the contribution of newer technologies from fixed telephony to ADSL and Fibre-To-The-Cabinet (FTTC) broadband (Ford and Koutsky, 2005; Koutroumpis, 2009; Qiang, Rossotto and Kimura, 2009; Czernich et al. 2011). Concerns have been expressed that inadequate provision of broadband infrastructure may severely disadvantage certain regions and firms, particularly in rural areas, leading to new digital inequalities (Riddlesden and Singleton, 2014; Oughton, Tyler and Alderson. 2015; Hilbert, 2016). Disparities in broadband often emerge due to market failure driven by (i) low adoption rates affecting demand-side economic viability (Manlove and Whitacre, 2019; Kongaut and Bohlin, 2016), (ii) economies of scale affecting supply-side economic viability often due to population density (Oughton et al. 2018) and (iii) sub-optimal regulatory practices (Bauer, 2010; Cave, Genakos and Valletti, 2019).

How broadband affects economic development can be measured in several ways (productivity, income, employment, innovation, poverty reduction, business creation etc). Most studies in the literature focus on the contribution of broadband to economic growth. For example, Holt and Jamison (2009) review studies on the relationship between broadband and economic growth in the US. Although in general there appears to be a positive relationship, methodological uncertainties and a lack of fine-grained geographical data prevent more definitive conclusions from being drawn. The results of Whitacre, Gallardo and Strover (2014) suggest broadband adoption significantly contributes to economic performance in rural US areas, but it is those non-metropolitan areas with high broadband adoption that display the largest rates of economic growth and reductions in unemployment. By contrast, the non-metropolitan areas with low adoption rates experience a decline in both firm presence and employment, suggesting that the spatial impacts are more nuanced.

Several studies explore the relationship between broadband and economic growth beyond the US. Koutroumpis (2009) investigates the effect of broadband on 22 OECD countries between 2002 and 2007, using a simultaneous equation model. The results suggest that the average effect of broadband on gross national product to be 0.4% per annum. Approximately 9.8% of the total growth over the study period could be attributed to broadband. More recent analysis continues to support this conclusion (Koutroumpis, 2019). Czernich et al. (2011) conduct an extensive study on the effect of broadband penetration on economic performance across OECD countries from 1996 to 2007. To minimise the potential problem of endogeneity, instrumental variables are used to estimate broadband. After controlling for country and year fixed effects, the analysis estimates that a 10% increase in broadband penetration leads to a 0.9-1.5% increase in the rate of per capita economic growth. However, local human capital is essential for broadband to be translated into local productivity gains (Mack and Faggian, 2013), and such gains may differ by industrial sector. For example, in manufacturing firms evidence suggests there is little benefit from broadband adoption (Haller and Lyons, 2015) demonstrating heterogeneous economic impacts.

Another major area of study is the potential for broadband to generate additional employment. Kolko (2012) finds that US employment was positively affected by broadband availability between 1999 and 2006. The effects are particularly strong in industries which rely more on information technology, and in areas which are less densely populated. A later study by Jayakar and Park (2013) reports similar findings, with a 10% increase in broadband availability reducing the unemployment rate by 0.7%. However, when Kolko (2012) examines metrics such as unemployment and the average wage, there appears to be little if any effect which can be attributed to broadband. This raises the important question of whether broadband truly brings benefits to local residents, with the result implying that the actual advantages for local economic development being rather limited. This is a key reason why there is still an ongoing debate about the economic impacts of broadband. Local analysis suggests that evidence associating job growth with broadband availability is limited (Ford and Seals, 2019), despite the recent introduction of policies to build municipal broadband networks. Indeed, examination of the relationship between faster broadband speeds and better economic outcomes suggests little economic payoff, with questions raised about the methods used in broadband assessments which tend to assume very large beneficial effects, such as with employment gains (Ford, 2018).

With respect to the potential economic geography impacts of broadband, the literature identifies three different schools of thought regarding firm location – the de-concentration school, the concentration school, and the heterogenous effect school (Mack and Grubesic, 2009; Mack, Anselin and Grubesic, 2011; Mack, 2014). The de-concentration school believes that geographic distance will largely disappear over time as broadband enables firms to locate further from central locations without incurring the disadvantages associated with distance. Some believe that broadband can act as a substitute to face-to-face interactions, thus negating the need to meet in person (Salomon, 1996, 1998; Kolko, 2012). The concentration school on the other hand believes that the presence of broadband will reinforce the advantages of central urban locations and that digital communications cannot replace the role of face-to-face interaction (Glaeser, 1998; Kolko, 2010). For example, broadband may lower transaction costs increasing spill-over effects, thus encouraging agglomeration in the manner described by Glaeser and Kohlhase (2004). In contrast, the heterogenous school combines elements of both perspectives, whereby broadband does not by itself result in decentralisation but provides the option for certain industries to locate in non-central locations. Therefore, whether broadband ultimately results in decentralisation is highly dependent on industry specific preferences and locations (Atkinson, 1998; Moss, 1998; Audirac, 2005; Forman, Goldfarb & Greenstein, 2005).

The influence of ICT on business location varies according to the firm business model adopted. New companies may require the benefits of urban locations, whereas mature firms with established business models (and brand reputation) may be satisfied with cheaper, less central locations (Campi, Blasco and Marsal, 2004). There is a lack of empirical evidence to verify the de-concentration school which *a priori* presumes that sufficient broadband infrastructure is available in rural and remote locations. Yet, due to heavy capital costs roll-out is not guaranteed, hence broadband displays a strong urban bias (Grubesic and Murray, 2002, 2004; Grubesic, 2006, 2008; Mack and Grubesic, 2009).

Mack et al. (2011) conduct a study on the effect of broadband provision on knowledge intensive business location in selected US metropolitan areas. Stronger effects of broadband on business location can be observed when the models are analysed at the metropolitan level compared to the national model where data are aggregated. Parajuli and Haynes (2012) conduct a study on Ohio and Florida using cross-sectional data in 2006. Some of the earlier empirical studies employ ordinary least square (OLS) analysis, which suffers from several major flaws. Firstly, it is shown that both broadband provision (Grubesic, 2006, 2008) and firm location (Carroll, Reid and Smith. 2008; Banasick, Lin, and Hanham, 2009; Mack & Grubesic) exhibit spatial autocorrelation. Secondly, reverse causality can run from business creation to broadband provision, leading to an endogenous relationship which must therefore be controlled for (Tranos and Mack, 2016).

Many empirical studies document the phenomenon of broadband and business creation linkages, mainly in the US context. For example, Mack and Grubesic (2009) conduct a study on the state of Ohio, using data from 1999 to 2004. Aggregate state results fail to demonstrate any significant relationship. Only when they disaggregate the data by industrial sector the relationship becomes observable, which again implies a more nuanced relationship than the conventional narrative. Firm size also significantly impacts the relative importance of broadband infrastructure, where small firms are likely to be the main beneficiaries compared to medium and large firms. The OLS model and the spatially weighted models exhibit significantly different results, with the latter displaying a better fit at both the aggregate and industrial sector level. This suggests that the relationship between broadband and business creation displays substantial spatial variations.

Aside from quantitatively estimating the relationship between broadband and business creation, several studies highlight multidimensional patterns supporting the predictions of the heterogenous effects school. For example, Mack (2014) finds the relationship between broadband and firm location to be non-uniform, localised and context specific. In some cases, broadband provides a crucial link for businesses to locate in low cost regions yet maintaining connectivity with markets and knowledge centres, while in other areas such benefits are minimal. Mack and Rey (2014) examine more specifically the variations in the relationship between broadband and knowledge intensive businesses across different metropolitan areas. It is revealed that out of the 54 metropolitan areas studied, only five do not show statistically significant results. The study also delves into the effect on this relationship attributable to the size of the metropolitan area and its industrial legacy. Traditional industrial states such as Ohio, Indiana and Pennsylvania display comparatively low coefficients for broadband's impact on business establishment creation. This indicates that they benefit relatively little from broadband roll-out and may be placed at a relative disadvantage compared to regions which have become more de-industrialised, specialising in services.

Other studies focus on less common subsets of conditions. Li et al. (2016) study the location decisions of rapidly growing businesses, defined as private sector firms with significant increases in revenues which are not part of a subsidiary. The results show that in metropolitan counties with large existing firms, there is a high rate of establishment growth for fast growing businesses. This suggests significant clustering effects which benefit businesses in the digital sector. Kim and Orazem (2016) examine the effect of broadband on new firm entry in rural areas in the US. Rural areas are traditionally regarded as disadvantaged regions with suboptimal infrastructure and deficient human capital. The study shows that the effect of broadband is strongest in those rural areas which border populous urban economies, as broadband facilitates connections to the wealth-generating metropolitan areas. The rural industries which appear to benefit most from broadband are health and education services, while manufacturing is the least impacted.

A recent study by McCoy et al. (2018) explores the relationship between broadband and business creation in Ireland. This study reports a positive relationship between broadband and business creation between 2002 and 2011, after controlling for other factors (e.g. road network expansion). It also finds that there is an interaction with human capital where the effects of broadband on business creation becomes amplified in areas with a high proportion of the workforce holding tertiary university-level degrees. This interaction is true for all firms, but especially for high-tech firms.

Given the evidence evaluated in this review, we can conclude that the effect of broadband on business creation remains relatively unexplored compared to other types of economic impacts, such as growth or employment. Hence, there is strong motivation for focusing on this factor, given business creation is an important indicator of a region's economy, and a predictor of future economic growth (Carree and Thurik, 2010; Fritsch, 2011). Across the different assessments carried out, broadband generally has positive economic impacts in urban areas, but with mixed results in rural locations, as well as across industries (Forman, Goldfarb & Greenstein, 2005). This article adds to the relatively scarce existing literature specific to the relationship between broadband and business creation in three novel ways a) examining the relationship in England, United Kingdom – an economically important region outside of the US context that is of interest globally b) understanding the relationship at extremely fine geographical levels across space and time with a detailed panel dataset we have assembled and c) investigating NGA which is a relatively new form of digital infrastructure on business creation. This provides insight for the deployment of even newer generations of digital technology which governments across the world are considering, such as 5G.

3. METHODOLOGY

The method focuses on using local area statistical units ('Middle Super Output Areas') in England (N=6,791), roughly equivalent to towns and municipalities in the US division, and NUTS IV (Nomenclature of Territorial Units for Statistics) in Europe. As far as we are aware McCoy et al. (2018) is the only study which uses data of similar granularity for studying the impacts of broadband on business creation, via 858 Irish Electoral Divisions (ED) aggregated in 192 'Urban Fields'. The aggregation which ended up significantly reducing their granularity is necessary in the case of McCoy et al. (2018) since many EDs have no firms, which would result in the 'severe excess zero problem'. This is not an issue for our study since England is substantially more populous than the Republic of Ireland, and all statistical areas contain business establishment units. Therefore, unlike McCoy et al. (2018) we can afford to use an even higher geographical resolution without spatial aggregation. As defined in various studies, an establishment is an operative local unit of a business such as a branch, factory or shop (Mack and Grubesic, 2009; Mack, Anselin and Grubesic, 2011; Mack, 2014; Mack and Rey, 2014; McCoy et al. 2018), which is a more accurate way of capturing local business operations, rather than using the head office location.

We estimate the econometric model specified in Equation (1):

(1) $$\frac{Y_{it+1}-Y_{it}}{Y_{it}} = \alpha + \beta I_{it} + \gamma X_{it} + \mu_i + \delta_t + \varepsilon_{it}$$

Where $Y_{it}$ denotes the number of business establishments in area $i$ at time $t$, hence the dependent variable denotes the *rate* of change. Using the rate variable helps to capture the proportion of change in a given area and hence reflect the number of existing establishments during the base year. $I_{it}$ is the indicator of broadband infrastructure and $\beta$ the coefficient, $X_{it}$ a matrix of control variables with a vector of coefficients denoted by $\gamma$. $\alpha$ represents the intercept, $\mu_i$ is the time invariant characteristic unique to each area, $\delta_t$ denotes time fixed-effect, which can be captured by a list of time dummies and $\varepsilon_{it}$ is the error term.

Equation (1) is widely used in similar studies (e.g. Mack, Anselin and Grubesic, 2011; Mack & Rey, 2014; McCoy et al. 2018). However, we do not investigate the elasticity relationship but rather the relationships in level form, given that our main dependent and independent variables of interest are both measured in percentages, as will be explained later in the data section.[1] Moreover, rather than modelling the stock of firms as the dependent variable, we examine the change in the business establishment growth rate, which is a flow rather than stock variable and should be less vulnerable to endogeneity problems. This metric should also be less persistent temporally than a business stock variable, which may pose problems for the stationarity of the data. The stock of local establishments is also highly dependent on local factors and characteristics such as history, especially in a country such as England with an old industrial base, for which we cannot necessarily acquire complete local historical data. Although the process of a fixed-effects model removes time *invariant* heterogeneity, it does not guarantee the removal of time *varying* heterogeneity (or any heterogeneity interacting with components in the error term which are time varying). This is another justification for modelling the dependent variable as the *rate* of change because it is likely to be less temporally persistent than the stock variable of business establishments.

---

[1] The independent variable is the percentage of premises covered by Next Generation Access (NGA) broadband, bounded between 0 to 100%. The dependent variable however is not bounded and can be either positive or negative, hence it is valid to use linear models which assume a random distribution of the dependent variable.

For the purpose of this study, we are interested in finding the magnitude of the causal effect, $\beta$ in Equation (1). The fixed-effects model is strongly preferred to the random-effects model among economics and social sciences in general (Wooldridge, 2010). This is mainly due to model properties which eliminate the effect of time invariant unobservables that could also be correlated with the broadband variable. Random effects models on the other hand requires a stronger assumption of unobservables being uncorrelated with the broadband variable, which in this case can be unrealistic.

There is the possibility that major metropolitan areas, especially London, would exert strong effects which could bias the estimates for the entirety of England. We therefore analysed London separately, as well as excluded the city from the sample, like the approach taken by McCoy et al. (2018) in excluding Dublin for the Republic of Ireland.

Spatial dependency is also a potential problem which can result in biased estimates. We tested for spatial dependency using the Moran's I test. In addition, we computed a spatially weighted measure of broadband coverage in neighbouring areas. This helped to account for the spatial effect of broadband in neighbouring MSOAs. Equation (1) is modified as follows:

(2) $\quad \frac{Y_{it+1} - Y_{it}}{Y_{it}} = \alpha + \beta I_{it} + \gamma X_{it} + \theta W I_{it} + \mu_i + \delta_t + \varepsilon_{it}$

Where $W$ is the spatial weighting matrix. The spatial effect of broadband infrastructure on business creation is captured by the coefficient $\theta$. Typically, in the infrastructure context $\theta$ is negative, which demonstrates the economic displacement effect on neighbouring areas. The displacement effect occurs as infrastructure attracts mobile factors of production from other areas thus reducing the net economic gains, which is often the case when modelling transport infrastructure. However, it is possible for $\theta$ to be positive in which case it can be interpretated as a positive spillover or network effect from infrastructure. We make no assumption in Equation (2) regarding the sign of $\theta$.

Equation (2) presents a specification like that employed by Boarnet (1998) who modelled the displacement/spillover effect of infrastructure across various dimensions. The most common spatial weight matrices for physical proximity are the contiguity and distant decay matrix. We employed the former which is significantly less computationally demanding. Given that our purpose was only to account for, rather than to produce, a quantitatively accurate estimate of spatial dependence, this simpler method should suffice. For contiguity matrices geographically adjacent areas are designated with a value of 1 while all others assigned 0.

*Accounting for endogeneity*

Any quantitatively estimated relationship between broadband and business creation may be subjected to endogeneity bias, which arises chiefly from possible reverse causality and confounding variables bias. We therefore employed a Two-Staged-Least-Squared (2SLS) model, selecting an appropriate Instrumental Variable (IV). This forms a structural model whereby the endogenous broadband variable is set as a function of all exogenous variables, including the chosen IV. The first stage of this structural equation is specified by Equation (3):

(3) $\quad I_{it} = \alpha + \beta_{iv} IV_{it} + \gamma_1 X_{it} + \mu_i + \delta_t + \varepsilon_{it}$

Using Equation (3) we produce an estimate of exogenised broadband infrastructure $\hat{I}_{it}$ variable, which is substituted into Equation (1) in place of $I_{it}$.

The instrumental variable of choice is based on the roll-out data for Asymmetric Digital Subscriber Line (ADSL) technology by British Telecom (BT) between 1999-2007 available via SamKnows (2020). Telephone exchange locations and serving postcodes are identified using a postcode dataset available from the Office of National Statistics Open Geography Portal. Ordnance Survey Codepoint data are then obtained from Edina Digimap (Ordnance Survey, 2020) for the number of premises per postcode and allocated an ADSL-enabled date depending on when the telephone exchange was upgraded. Data are then aggregated to the MSOA, based on a postcode to MSOA lookup table available from ONS Open Geography Portal. Finally, the percentage coverage of premises in each MSOA for each year over the study period was calculated.

The approach of using historical and non-directly related infrastructure metrics as IVs for potentially endogenous infrastructure variables has been applied by Duranton and Turner (2012) in the US context, and later by Möller and Zierer (2018) in the German context, using similar justification. Historical infrastructure data are unlikely to be affected by reverse causality or confounding variables while still being correlated with current infrastructure variables, making them suitable as instruments in this context. Using data on the roll-out of ADSL2+ technology from 2001 to 2005 at the MSOA level we instrumented the broadband variable for 2011 to 2015. This period is appropriate given that most of the roll-out across England happened during this period. With a decade lag we can control for reverse causality, given that current business creation could not have affected the roll-out of ADSL 10 years ago. This removes a major source of endogeneity bias.

The roll-out of ADSL2+ broadband is the previous generation of broadband technology before FTTC. Its deployment would have also been subjected to urban bias. Generally, the exogenous factors which affected infrastructure deployment (ease of roll-out, deployment costs, topography and existing infrastructure and planning networks) are unlikely to have changed much over time and are reflected in our chosen IV. Hence, the lagged ADSL roll-out data is a strong predictor of current NGA roll-out, demonstrating relevance with the endogenous variable. Statistically this can be demonstrated by their high correlation (>0.6). Moreover, there are generally no unaccounted variables which are likely to affect lagged values of ADSL (and in turn NGA broadband), except for past population size. This means that the 2SLS model is unlikely to introduce additional endogeneity via the IV of lagged ADSL values, since other factors impacting lagged ADSL values are also unlikely to impact NGA. In conclusion, lagged values of ADSL are a strong IV for NGA broadband data, thus helping to account for potential endogeneity.

A potential weakness of lagged infrastructure variables as identified by both Duranton and Turner (2012) and Moller and Zeirer (2018), is that they are correlated with past population values which would have influenced roll-out decisions. Past values of population may also have influence, or are at least strongly correlated with current infrastructure, thus introducing a possible source of additional endogeneity. In order to limit this, we include lagged values of population for local statistical areas over 10 years, as suggested by Duranton and Turner (2012). Further statistical demonstration of the suitability of our IV is reported in Table 3.

4. DATA

A data summary is shown in TABLE A1. Most variables selected are referenced from existing studies in the literature, particularly from McCoy et al. (2018). The dependent variable in these models is the rate of change in the number of business establishments for the local statistical areas, which can be accessed via ONS Nomis (2018) (see Table A1). We break business establishments down by industrial sectors as classified by the Standard Industrial Classification of Economic Activities (SIC, 2007). Broad industrial groups are used which divide business establishments into 18 industrial sectors. The full list is shown in TABLE A4. However, due to strong geographic concentration in business establishments by sector, using 18 sectors would result in the excess zero problem where a substantial number of local statistical areas have no establishments. Thus, we amalgamated data into 8 industrial groupings which are listed in TABLE A5. FIGURE 1 shows geographically the percentage change in business establishments during the study period. It should be noted that our data denotes *net* percentage change in the number of business establishments. We used this since the data for establishment entry, exit and survival are not available in England at the MSOA level. Moreover, using net change has the advantage of reflecting the area's ability to both attract new entrants, as well as to discourage existing businesses from relocating or cease operating.

We included broadband (NGA) coverage as the independent variable of interest. Broadband data are taken from Ofcom's Infrastructure Reports (2012, 2013, and 2014) and the Connected Nations reports (2015, 2016), providing time-series data on the percentage of premises by postcode with NGA. This provides us with annual data from 2011 to 2015. By merging each postcode with premises data from Ordnance Survey's Codepoint database, the number of premises in each postcode with access to NGA could be estimated. After aggregating postcode data into local statistical areas using an ONS lookup table, the number of premises with NGA is the numerator, and the total number of premises is the denominator. The percentage of premises can then be obtained for each local statistical area with higher capacity broadband access for each time step. Those premises without NGA are generally on legacy ADSL2+ (up to 24 Mbps) with only a few very rural exchanges on ADSL (up to 8 Mbps). The percentage change in broadband coverage over the study period is shown in FIGURE 2.

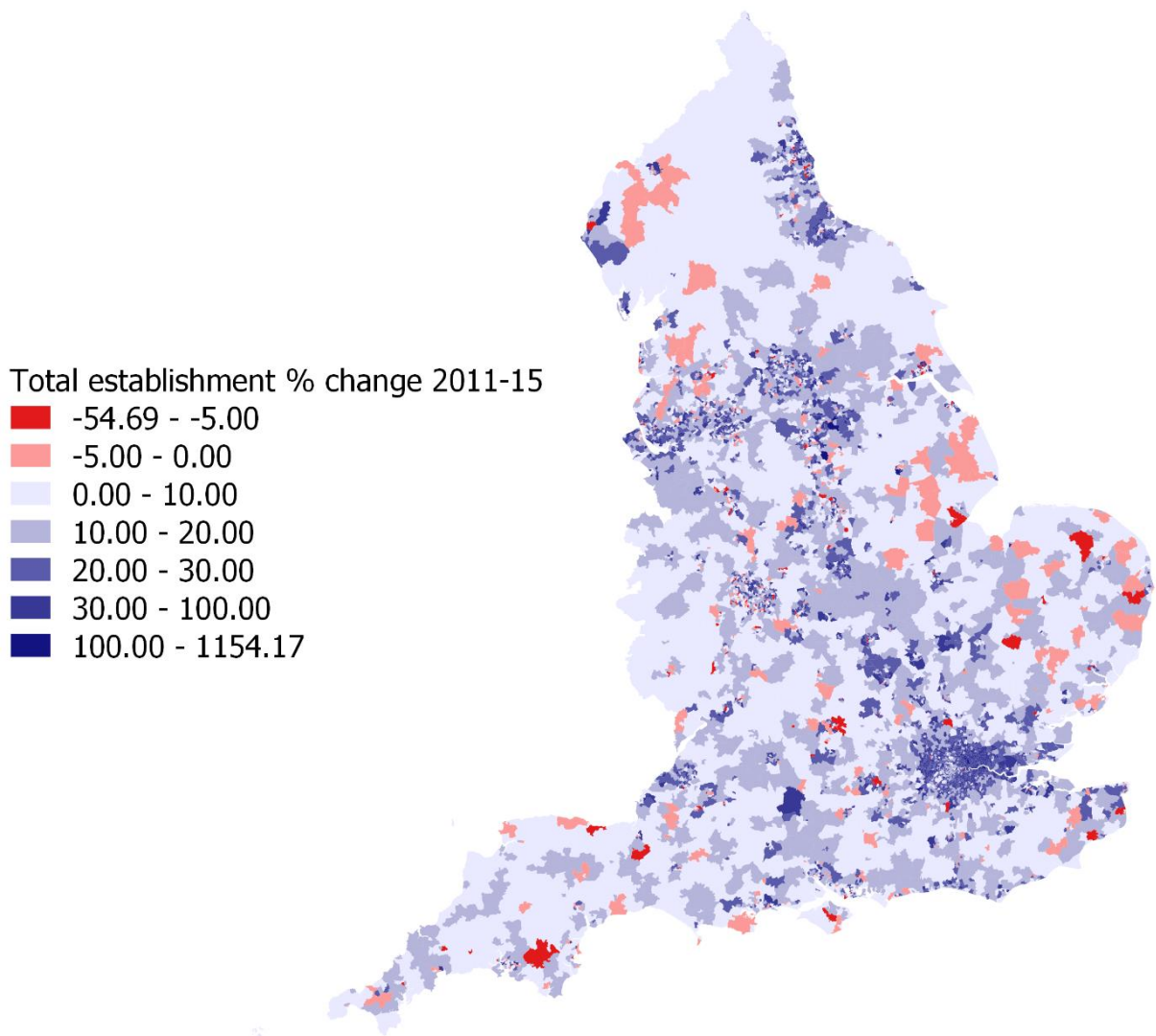

FIGURE 1: Total Establishment Percentage Change Over the Study Period (2011-2015)

We included a measure of transport infrastructure as it is frequently included as a control variable in previous studies which examine the impact of ICT and broadband (Mack, Anselin and Grubesic, 2011; Kolko, 2012; Mack and Rey, 2014; McCoy et al. 2018). The variable is constructed from traffic count data published by the Department for Transport (DfT, 2018). The DfT establishes count points on all major roads, important sections and junctions, as well as minor roads. Traffic flow for all times of day across the year are closely monitored dynamically and vehicle counts are identified by types. The count points also contain information regarding the road category on which the measurements are taken, as well as the length of the road segment. We then summed the length of all road transport routes falling within the boundary of each local statistical area, using GIS software. The result was then divided by the area size to obtain the road density per square kilometre.

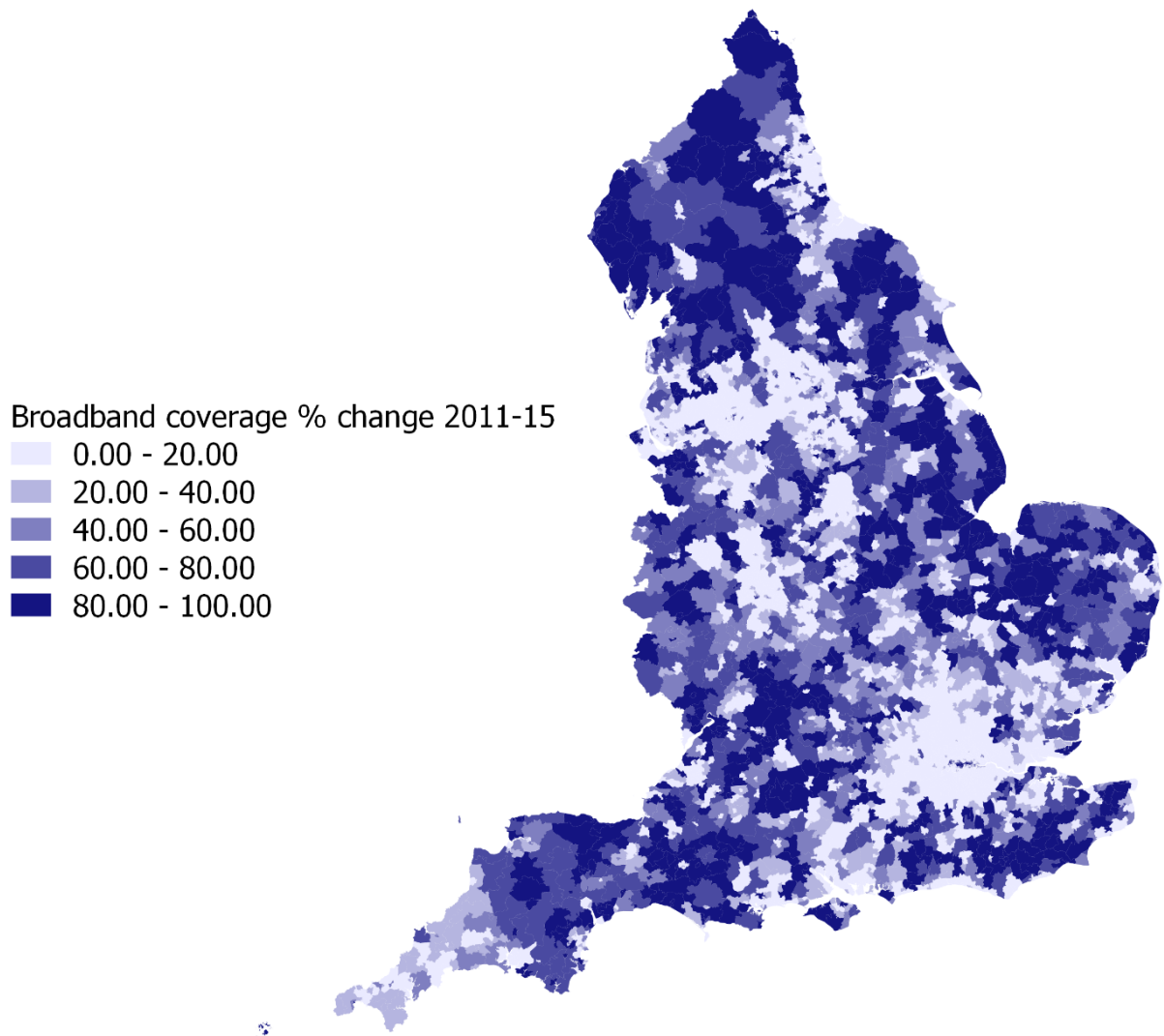

FIGURE 2: Broadband Coverage Percentage Change Over the Study Period (2011-2015).

Given the fine geographic area, there is likely to be greater local heterogeneity compared to more aggregated geographical units where extreme values are averaged out. We therefore included several other control variables to reflect the characteristics of the local area. They included the proportion of workforce with tertiary degree or above capturing the stock of local human capital; the inverse travel time to supermarkets; the rate of crime and antisocial behaviour (proxies for local amenities and area attractiveness, respectively); the inverse travel time to the nearest town centre and employment hub (proxying for settlement centrality and access to the nearest labour pool, respectively); an index of labour cost; a Herfindahl-Hirschman (HH) index measuring the degree of sectoral agglomeration or specialisation; population density which reflects the degree of urban agglomeration economies and the unemployment rate to capture the level of demand and factor usage, respectively. These variables were chosen as they have been commonly used in various studies (e.g. Kolko, 2012; Mack et al. 2011; Mack and Rey, 2014).

The proportion of the workforce with tertiary degree or above is an indication of the human capital or reflection of local labour force quality. It was obtained from the 2011 census (Census, 2011). Unfortunately, for local statistical areas this variable exists only for the 2011 census year. Therefore, rather than including it as a variable in the main regression, we used it to separate local statistical areas into low, medium and high human capital locations. The three categories are defined as areas with less than 35%, 35-45% and more than 45% of workforce with tertiary degrees or above, respectively. This numeric division was selected as it separated the total number of local statistical areas in England into three groups roughly equal in number.

The inverse travel times to town centres, supermarkets and employment centres were obtained from the UK DfT accessibility index which is based on the travel time by foot or public transport (DfT, 2011; 2018). Crime rate and antisocial behaviour data were obtained from the UK police websites (DATA.POLICE, 2020). The labour cost index was constructed using the UK Office for National Statistics (ONS) quarterly estimates of labour cost divided by industrial sectors, seasonally adjusted. We averaged the quarterly estimates to produce annual figures. We then used the proportion of employment in each sector for local statistical areas to compute a weighted index of the labour cost. The division of labour cost and employment proportion by sector were based on the broad industrial group list. Data for the population density and unemployment rate were obtained from the ONS.

The Herfindahl-Hirschman (HH) index of sectoral employment concentration was computed using the following formula:

(4) $\quad HH_{it} = \sum_{j=1}^{J} s_{ijt}^2$

Where $j = 1, 2, \ldots J$ denotes each industrial sector, while $s_{ijt}$ refers to the share of employment of sector $j$ in area $i$ at time $t$. The division of employment by sector was also based on ONS broad industrial groups.

5. RESULTS AND ANALYSIS

This section reports the regression results of the econometric analysis on Equations (1) to (3). We applied the cross sectional and time fixed-effects estimator for all our models. All standard errors are cluster robust and reported in parentheses. Table 1 summarises our main results. We test for spatial dependency using Moran's I which is reported in Table 2. Table 3 shows the first-stage regression result of our 2SLS model. In Table 4 we break down our regression results by several industrial sector groupings.

The main results are shown in Table 1 with 8 different fixed-effects specifications. Model 1 is the most parsimonious OLS specification controlling for a basic set of covariate'. We consider this our default estimate and reference point to compare all other models. The result indicates that a 1% point increase in NGA coverage results in 0.0185% percentage point *decrease* in the *rate* of business creation at the local level. This coefficient is statistically significant at the 1% level.

Models 2 to 6 are estimated using sub-samples of geographical areas. Model 2, Model 3 and Model 4 show the results for local statistical areas defined as having low (<35%), medium (35-45%) and high (>45%) human capital respectively in terms of the share of the workforce with tertiary degrees or above. The direction of the relationship between broadband and the rate of business creation is negative in all three instances but only statistically significant for medium and high human capital areas. The coefficients for medium and high human capital areas are -0.0136 and -0.0210, respectively. For Model 5 we examine specifically the London region. There does not appear to be a significant relationship between broadband and rate of change in business creation in London. In Model 6 we exclude London from our analysis, and the model shows a negative relationship which is statistically significant. Apart from the removal of London, Model 6 and Model 1 are otherwise identical and likewise the size of the coefficients is similar at -0.0161 and -0.0185, respectively, which indicates that our default result in Model 1 is not substantially affected by the inclusion of London.

In Model 7 we apply the spatial model of the form shown in Equation (3). After controlling for the weighted NGA coverage of adjacent areas, there is no statistically significant relationship between broadband and the rate of change in business creation. The weighted broadband variable however is negative and statistically significant at 1%, which suggests a small negative displacement effect of broadband on business creation. Table 2 displays the results of Moran's I test for spatial dependency. Despite very large positive spatial correlation showing clustered values, none of the metrics are statistically significant, indicating that spatial autocorrelation is not a major issue in this analysis. Although the dependent variable does experience an increase in spatial dependency from the base year 2011 to the final year 2015, our spatial model is able to substantially reduce the spatial dependency of the residuals as shown in the last column of Table 2. In Model 8 we apply the 2SLS model to remove the potential endogeneity between NGA coverage and growth rate of business establishments. The result indicates that a 1% point increase in NGA coverage leads to 0.0314% percentage points *decrease* in the *rate* of business creation at the local level. In Table 3 we show the first stage regression. The signs of the variables mostly conform to common expectations, which suggests that our models appear to be theoretically sound. For example, there is positive and significant association of business establishment change with road density, inverse travel time to town centre, population density and areas with high labour costs, which reflects the urban bias of NGA roll-out. Our selected IV variable is the coverage of ADSL broadband between years 2001 and 2005 (10-year lag). It is also positively related to NGA and highly statistically significant with a t-value of over 42. A 1% point increase in ADSL coverage ten years prior is associated with a 0.316% percentage points increase in NGA coverage. The rule of thumb is that for a single t-statistic or the joint F-statistic for multiple IVs should exceed 10 (Stock and Yogo 2005). The extra control variable with 10-year lagged population is also positively associated with current NGA coverage, as would be expected. These combined with a high correlation of 0.62 between our IV and endogenous variable demonstrates a high degree of relevance for the IV we have selected.

In Table 4 we divide firms into industrial sector classifications as shown in Table A5. In all the sectors there appear to be a negative association but the coefficients are only significant for four sectors – "Manufacturing and construction", "Wholesale, transport and storage", "Public sector and health" and "Primary industries and utilities".

*Table 1. Regression results of growth rate of business establishment against the independent variables*

| Dependent variable: growth rate of business establishment | Model 1 | Model 2 | Model 3 | Model 4 | Model 5 | Model 6 | Model 7 | Model 8 |
|---|---|---|---|---|---|---|---|---|
| *Broadband premises percentage coverage* | -0.0185*** | -0.0105 | -0.0136*** | -0.0210*** | -0.0389 | -0.0161*** | -0.00418 | -0.0314*** |
| | (0.00385) | (0.00817) | (0.00440) | (0.00487) | (0.0237) | (0.00397) | (0.00474) | (0.00856) |
| Road density (km/hectare) | -0.00354 | 0.00290 | -0.00241 | -0.0144 | -0.00525 | -0.00277 | -0.00309 | -0.00302 |
| | (0.00643) | (0.0118) | (0.00527) | (0.0126) | (0.0119) | (0.00694) | (0.00647) | (0.00639) |
| Inverse travel time to nearest town centre | -0.0115** | -0.00934 | -0.0146** | -0.00723 | -0.00130 | -0.0130*** | -0.0105** | -0.0104** |
| | (0.00451) | (0.00835) | (0.00621) | (0.00746) | (0.0154) | (0.00477) | (0.00451) | (0.00464) |
| Inverse travel time to nearest employment centre | 0.0196*** | 0.0131* | 0.0179*** | 0.0215*** | 0.0188** | 0.0184*** | 0.0181*** | 0.0140*** |
| | (0.00314) | (0.00702) | (0.00269) | (0.00407) | (0.00947) | (0.00328) | (0.00324) | (0.00341) |
| Inverse travel time to nearest supermarket | -0.00494 | -0.00201 | 0.0225*** | -0.00867 | -0.0270* | 0.00125 | -0.00890 | -0.00615 |
| | (0.00594) | (0.0120) | (0.00645) | (0.00995) | (0.0150) | (0.00619) | (0.00616) | (0.00614) |
| Crime and anti-social behaviour per 1000 | 0.00260 | 0.0104 | -0.00454 | -0.0138** | -0.0208 | 0.00801* | 0.00228 | 0.00153 |
| | (0.00435) | (0.00899) | (0.00611) | (0.00623) | (0.0151) | (0.00454) | (0.00435) | (0.00437) |
| Labour cost index | -0.0765 | -0.102 | -0.208* | 0.0653 | 0.153 | -0.0838 | -0.0733 | -0.0678 |
| | (0.0582) | (0.0782) | (0.108) | (0.142) | (0.272) | (0.0597) | (0.0582) | (0.0563) |
| Herfindahl-Hirschman specialisation index | -0.0150** | -0.00778 | -0.0154 | -0.0270** | -0.0146 | -0.0146* | -0.0153** | -0.0143** |
| | (0.00684) | (0.0118) | (0.00987) | (0.0112) | (0.0120) | (0.00777) | (0.00683) | (0.00676) |
| Population density | -0.120** | 0.00764 | 0.118** | -0.106 | -0.113 | -0.155*** | -0.123** | -0.0294 |
| | (0.0557) | (0.0672) | (0.0600) | (0.101) | (0.154) | (0.0481) | (0.0557) | (0.0567) |
| Unemployment rate | 0.0376 | -0.484 | 0.844 | 2.372*** | -0.962* | 0.165 | 0.109 | -0.0651 |
| | (0.255) | (0.365) | (0.544) | (0.451) | (0.556) | (0.271) | (0.253) | (0.240) |
| Weighted broadband coverage of adjacent MSOAs | - | - | - | - | - | - | -0.00516*** | - |
| | | | | | | | (0.00108) | |
| Population 10 year lag | - | - | - | - | - | - | - | -0.294*** |
| | | | | | | | | (0.0351) |
| Intercept | 0.769** | 0.601 | 1.010 | -0.219 | -0.390 | 0.860** | 0.761** | 2.408*** |
| | (0.388) | (0.533) | (0.707) | (0.875) | (1.530) | (0.398) | (0.388) | (0.336) |
| | (0.00558) | (0.00885) | (0.0111) | (0.0117) | (0.0221) | (0.00588) | (0.00560) | (0.00558) |
| N: | 33955 | 13405 | 10700 | 9850 | 4915 | 29040 | 33955 | 33955 |
| Groups: | 6791 | 2681 | 2140 | 1970 | 983 | 5808 | 6791 | 6791 |

*, ** and *** denote statistical significance at the 10%, 5% and 1% level, respectively

*Table 2. Moran's I test for spatial dependency*

| Year | Business establishment growth rate | Model residual | Model residual after accounting for neighbouring infrastructure |
|---|---|---|---|
| 2011 | 0.027 | 0.540 | 0.485 |
| 2015 | 0.146 | 0.315 | 0.288 |

*, ** and *** denote statistical significance at the 10%, 5% and 1% level, respectively (null hypothesis: no spatial dependency)

*Table 3. First stage regression and diagnostic tests on Model 8*

| Dependent variable: broadband premises percentage coverage | First stage | |
|---|---|---|
| ADSL percentage coverage 2001-2005 (instrumental variable) | 0.00316*** | |
| | (0.0000746) | |
| Road density (km/hectare) | 0.0177* | |
| | (0.00961) | |
| Inverse travel time to nearest town centre | 0.135*** | |
| | (0.0134) | |
| Inverse travel time to nearest employment centre | -0.0472*** | |
| | (0.00543) | |
| Inverse travel time to nearest supermarket | -0.140*** | |
| | (0.0119) | IV t-statistic: 42.36*** |
| Crime and anti-social behaviour per 1000 | -0.00538 | IV endogenous variable correlation: 0.62*** |
| | (0.00924) | Durbin-Wu-Hausman (k=1,n=6790): 0.740 |
| Labour cost index | 0.170*** | (p=0.460) |
| | (0.0257) | |
| Herfindahl-Hirschman specialisation index | -0.0129*** | |
| | (0.00830) | |
| Population density | 0.320*** | |
| | (0.0733) | |
| Unemployment rate | -2.103*** | |
| | (0.168) | |
| Population 10 year lag (additional control variable due to instrumental variable) | 0.234*** | |
| | (0.0962) | |
| Intercept | -3.331*** | |
| | (0.779) | |
| N: | 33955 | |
| Groups: | 6791 | |

*, ** and *** denote statistical significance at the 10%, 5% and 1% level, respectively

*Table 4. Regression results of growth rate of business establishment against the independent variables, broken down by industrial sectors*

| Dependent variable: growth rate of business establishment | Knowledge intensive | Manufacturing & construction | Arts, entertainment, recreation & other services | Wholesale, transport & storage | Public sector & health | Retail & motor trades | Accommodation, food services & property | Primary industries and utilities |
|---|---|---|---|---|---|---|---|---|
| Broadband premises percentage coverage | 0.000670 | -0.0142** | -0.0103 | -0.0414*** | -0.0633*** | -0.00663 | 0.00194 | -0.0374*** |
| | (0.00566) | (0.00580) | (0.00845) | (0.00985) | (0.00940) | (0.00611) | (0.00859) | (0.0129) |
| Road density (km/hectare) | 0.00153 | -0.00297 | -0.00716 | -0.0328** | -0.0141 | -0.000856 | -0.00921 | -0.0113 |
| | (0.00973) | (0.0101) | (0.0114) | (0.0145) | (0.0131) | (0.00963) | (0.0129) | (0.0324) |
| Inverse travel time to nearest town centre | -0.00164 | -0.0113 | 0.00835 | -0.00605 | -0.0000384 | -0.00297 | -0.0318** | -0.0140 |
| | (0.00876) | (0.00911) | (0.0125) | (0.0185) | (0.0143) | (0.00925) | (0.0139) | (0.0219) |
| Inverse travel time to nearest employment centre | 0.0202*** | 0.0140*** | 0.00960* | 0.0254*** | 0.0254*** | 0.0108*** | 0.0108* | 0.00114 |
| | (0.00345) | (0.00401) | (0.00549) | (0.00763) | (0.00563) | (0.00408) | (0.00647) | (0.0100) |
| Inverse travel time to nearest supermarket | -0.0119 | 0.000368 | -0.0177 | -0.0204 | -0.00275 | 0.00595 | 0.0267** | -0.00156 |
| | (0.00817) | (0.00861) | (0.0112) | (0.0148) | (0.0118) | (0.00854) | (0.0136) | (0.0215) |
| Crime and anti-social behaviour per 1000 | 0.00705 | -0.00110 | 0.0101 | 0.0228 | 0.00210 | -0.00408 | 0.0173 | -0.0108 |
| | (0.00830) | (0.00897) | (0.0128) | (0.0172) | (0.0137) | (0.00974) | (0.0149) | (0.0230) |
| Labour cost index | 0.0309 | -0.260** | -0.309* | -0.397* | 0.501** | -0.0821 | 0.0749 | -0.0715 |
| | (0.123) | (0.117) | (0.187) | (0.238) | (0.200) | (0.142) | (0.249) | (0.352) |
| Herfindahl-Hirschman specialisation index | -0.0246** | -0.0234** | -0.0427*** | -0.0415** | -0.00899 | -0.0270** | -0.0255 | -0.0233 |
| | (0.0108) | (0.0114) | (0.0158) | (0.0201) | (0.0147) | (0.0131) | (0.0190) | (0.0248) |
| Population density | -0.118* | -0.101 | -0.224*** | -0.208** | -0.0449 | -0.370*** | -0.332*** | -0.275 |
| | (0.0670) | (0.0644) | (0.0790) | (0.104) | (0.0749) | (0.0722) | (0.0865) | (0.168) |
| Unemployment rate | 0.490 | 1.359*** | 1.715*** | 0.0217 | 0.843** | 1.625*** | 1.548*** | 2.742*** |
| | (0.397) | (0.374) | (0.404) | (0.487) | (0.389) | (0.398) | (0.465) | (0.669) |
| Intercept | 0.132 | 1.814** | 2.369** | 2.736* | -2.932** | 1.497* | 0.307 | 1.064 |
| | (0.783) | (0.732) | (1.175) | (1.497) | (1.244) | (0.896) | (1.552) | (2.224) |
| N: | 33955 | | | | | | | |
| Groups: | 6791 | | | | | | | |

*, ** and *** denote statistical significance at the 10%, 5% and 1% level, respectively

# 6. DISCUSSION AND CONCLUSION

In this article the empirical relationship between broadband and business creation was examined at the local level over the 2011-2015 period in England. After controlling for local area variables, our results show a *negative* relationship between broadband coverage and the *rate* of business creation. The negative relationship persists even after controlling for possible endogeneity with appropriate instruments, suggesting causational relationship. The negative relationship is also observed in areas with different levels of human capital, in London region and across all industrial sectors, though significant only for "Manufacturing and construction", "Wholesale, transport and storage", "Public sector and health" and "Primary industries and utilities". There is also evidence of a mild negative spatial displacement effect given the small but statistically significant coefficient of the weighted index of neighbouring broadband coverage. There are several possible explanations for the relationship observed in the data.

Firstly, it is possible that the provision of advanced digital infrastructures such as NGA broadband does not represent a major improvement over the existing available technology. Prior generations of ADSL enable download speeds of up to 24 Mbps, which is already sufficient for the vast majority of business or consumer day-to-day activities, which usually only requires speeds of 2-5 Mbps. The higher speeds enabled by NGA broadband (e.g. >10 Mbps) tend to satisfy many consumer utilities such as HD video streaming (e.g. Netflix) which are non-essential for most business processes. This view is consistent with authors such as Ford (2018) and Kenny and Kenny (2011) who argue that high-speed broadband's economic impact is over-stated and subsidies to support infrastructure delivery could waste billions of dollars of public money.

Secondly, it is widely acknowledged that broadband may set in motion Schumpeterian 'creative destruction' processes, with relatively unstudied impacts on the location of those newly created businesses, and indeed the location of those businesses that are destroyed. Particularly over the past decade there has been increased consolidation in many sectors, particularly in the digital economy. Often small firms trying to operate from a physical business unit have struggled to compete against larger online-only firms. Amazon is one example which personifies the shift from high street retail towards a much more consolidated marketplace. Thousands of specialised businesses with a physical high street presence now find it challenging to compete with the costs of production attained through large economies of scale and warehouse automation (and the ease with which customers can purchase from home). The negative relationship found in this analysis may therefore indicate that this ongoing process is potentially discouraging new businesses from starting in the first place due to this concentration of market power. Although not covered by our data, this trend is likely to be even further attenuated by the recent COVID-19 pandemic which led to a surge in online services at the expense of businesses who maintain a physical presence.

There are several limitations to this analysis which merit discussion. Firstly, the data analysed here covers a relatively short period of only five years, mostly during the period of NGA broadband roll-out. It is possible that the study period is too short to catch positive economic effects arising from NGA broadband. This is a well-known problem in economics related to ICT exhibiting a 'productivity paradox' where benefits may take decades to appear as they depend on there actually being 'killer applications' for the provided technologies. Secondly, it was beyond the scope of this analysis to explore whether the decrease in the rate of business creation, which could have been due to market consolidation, lead to increased employment, Gross Domestic Product or other beneficial economic impacts overall. Certainly, other studies which focus on these effects point to there being positive impacts (Koutroumpis, 2009; Czernich et al. 2011), but usually at the macroeconomic level and over previous time periods. It could be that the digital economy has now reached a different phase of development.

There are many policy ramifications to these findings. Firstly, there is almost always an assumption that faster broadband will lead to better economic outcomes, despite important contributions in the literature calling this assumption into doubt (Ford, 2018; Kenny and Kenny, 2011). Over the time period assessed in this analysis many governments around the world have pursued policies that aim to enhance the coverage and capacity of broadband infrastructure particularly in rural and remote regions. The findings of this article do not necessarily call for a divergence in aspiring to reach near-universal broadband connectivity, but that we should be much more realistic about the economic impacts in relation to the costs. Afterall, broadband is a necessary but not sufficient factor for economic development. While citizens may enjoy more online services (e.g. Amazon, Netflix etc.) and feel better able to participate in society, it could come at the cost of local businesses who lose out to cheaper online competitors, causing an erosion in the stock of local businesses with major repercussions for the local tax base. This study which employed high-resolution data is able to shed light on some of the dynamics at the very local level.

Considering the conclusions reached there are future areas of research which should be addressed to provide greater insight on this important issue. Further time series analysis must be undertaken in other countries to examine the degree to which these effects are applicable to other locations, ideally over a longer and more recent time horizon. One of the key contributions in this article was the high-resolution spatial data analysed, therefore replicating this approach in countries with this level of data granularity would be a valuable exercise, especially with an extended time series. Ideally future work would consider both the rate of change in new business creation as well as the rate of employment change, to provide greater insight on how broadband affects market consolidation and employment due to the use of online services.

## ACKNOWLEDGEMENTS

This work was supported by the UK Engineering and Physical Science Research Council programme grant entitled Multi-scale Infrastructure systems Analytics (EP/N017064/1). The authors would like to thank multiple anonymous reviewers for valuable feedback though the review process. There are no competing interests.

# APPENDIX

## TABLE A1: Summary of Data Used in Econometric Models

| Variable | Type | Description | Source |
|---|---|---|---|
| Growth rate of business establishment | Dependent variable | Number of business establishments, broken down by size and sector | ONS Nomis (2018): UK Business Counts |
| Broadband percentage coverage | Infrastructure variable | Percentage of premises in MSOA covered by NGA FTTC | Source: Ofcom (2012-2017) |
| Road density | Infrastructure variable | Length of major road per unit area. Major roads includes all principle and trunk roads as defined by the DfT. | DfT (2011, 2018): compiled from traffic count points |
| Motor vehicle density | Instrumental variable | AADF of motor vehicles per unit area | DfT (2011, 2018): compiled from traffic count points |
| Inverse travel time to town centre | Accessibility variable | The inverse travel time to the nearest town centre by walk or public transport. Denotes centrality. | DfT (2011, 2018): accessibility statistic, 2014 and 2015 adjusted for comparability with 2011-13 |
| Inverse travel time to supermarket | Accessibility variable | The inverse travel time to the nearest supermarket by walk or public transport. Proxy for local amenity and services. | DfT (2011, 2018): accessibility statistic, 2014 and 2015 adjusted for comparability with 2011-13 |
| Proportion of workforce with tertiary degree | Local characteristic | The percentage of workforce with higher education degree – an indication of the human capital or quality of local workforce. | Census for 2011 (Census, 2011) |
| Population density | Instrumental variable | Population per unit area | Local population estimates from ONS |
| Total employment | Instrumental variable | Total number of employment | ONS Nomis (2018): UK Business Register and Employment Survey |
| Unemployment rate | Instrumental variable | Unemployment rate by jobseekers' allowance. Reflects the level of local factor demand. | ONS Nomis (2018): UK Business Register and Employment Survey |
| Crime and antisocial behaviour per 1000 | Instrumental variable | Shows the crime rate in an area. Denotes the level of deprivation and poor environment, which should discourage business creation ceteris paribus. | DATAPOLICE (2020): UKCrimeStats, a data platform from the Economic Policy Centre |
| Labour cost | Local characteristic | Index of labour cost per hour. Another indication of local demand. Seasonally adjusted. | ONS Nomis (2018): ONS estimate of labour cost per hour seasonally adjusted, divided by industrial sector. Employment proportion in each sector used to compute local labour cost index. |
| HH index | Local characteristic | Spatial Herfindahl-Hirschman index measuring the degree of industrial specialisation. | Computed using local share of employment in each industry. |

## TABLE A2: Variable Descriptive Statistics 2011 (start year)

| Variable | Mean | Standard deviation | Minimum | Maximum |
|---|---|---|---|---|
| Business establishment growth rate | 0.0611% | 6.548% | -51.948% | 271.831% |
| Broadband premises percentage coverage | 76.671 | 33.511 | 0 | 100 |
| Road density (km/hectare) | $4.31 \times 10^{-6}$ | $6.22 \times 10^{-6}$ | 0 | $7.01 \times 10^{-5}$ |
| Motor vehicle density (traffic/hectare) | 0.0161 | 0.0352 | 0 | 0.608 |
| Inverse travel time to nearest town centre | 0.0823 | 0.0349 | 0.00833 | 0.2 |
| Inverse travel time to nearest supermarket | 0.155 | 0.0361 | 0.00833 | 0.2 |
| Percent of workforce with tertiary degree | 27.094 | 9.456 | 2.252 | 66.514 |
| Population density (persons/hectare) | 33.284 | 34.836 | 0.0573 | 247.732 |
| Total employment | 3541.678 | 8063.378 | 115 | 372730 |
| Unemployment rate | 0.0244 | 0.0171 | 0.00246 | 0.184 |

| Crime and anti-social behaviour per 1000 | 14,476 | 12.986 | 1.778 | 354.503 |
|---|---|---|---|---|
| Labour cost index | 465.086 | 16.847 | 350.375 | 488.212 |
| Herfindahl-Hirschman specialisation index | 0.184 | 0.0931 | 0.0782 | 0.837 |
| Population (not included in regression) | 7820.228 | 1611.838 | 2224 | 16439 |
| Area in hectares (not included in regression) | 1918 | 4766.451 | 23 | 111716 |
| ADSL percentage coverage in 2001 (IV 10-year lag) | 66.541 | 44.487 | 0 | 100 |
| Population in 2001 (10-year lag) | 7315.457 | 1407.807 | 113 | 15463 |

TABLE A3: Variable Descriptive Statistics 2015 (end year)

| Variable | Mean | Standard deviation | Minimum | Maximum |
|---|---|---|---|---|
| Business establishment growth rate | 3.977% | 9.443% | -90.435% | 181.657% |
| Broadband premises percentage coverage | 94.354 | 10.818 | 0.155 | 100 |
| Road density (km/hectare) | $4.377 \times 10^{-6}$ | $6.23 \times 10^{-6}$ | 0 | $7.01 \times 10^{-5}$ |
| Motor vehicle density (traffic/hectare) | 0.0161 | 0.0345 | 0 | 0.627 |
| Inverse travel time to nearest town centre | 0.0816 | 0.0359 | 0.00833 | 0.341 |
| Inverse travel time to nearest supermarket | 0.166 | 0.0655 | 0.00833 | 0.501 |
| Percent of workforce with tertiary degree | 30.588 | 10.512 | 3.324 | 85.454 |
| Population density (persons/hectare) | 34.626 | 36.907 | 0.0561 | 274.425 |
| Total employment | 3640.52 | 8001.99 | 170 | 439650 |
| Unemployment rate | 0.00892 | 0.00758 | 0.000413 | 0.0598 |
| Crime and anti-social behaviour per 1000 | 13.812 | 11.800 | 1.963 | 305.113 |
| Labour cost index | 503.672 | 16.674 | 383.073 | 531.216 |
| Herfindahl-Hirschman specialisation index | 0.179 | 0.0905 | 0.0748 | 0.864 |
| Population (not included in regression) | 8067.490 | 1788.104 | 2324 | 18438 |
| Area in hectares (not included in regression) | 1918 | 4766.451 | 23 | 111716 |
| ADSL percentage coverage in 2005 (IV 10-year lag) | 99.907 | 1.658 | 10 | 100 |
| Population in 2005 (10-year lag) | 7504.813 | 1464.996 | 1738 | 16336 |

TABLE A4: Industrial Sector Break Down According to Nomis Broad Industrial Sector Groups

| Label | Industrial group name |
|---|---|
| A | Agriculture, forestry & fishing |
| B, D and E | Mining, quarrying & utilities |
| C | Manufacturing |
| F | Construction |
| G | Motor trades |
| G | Wholesale |
| G | Retail |
| H | Transport & storage (including postal) |
| I | Accommodation & food services |
| J | Information & communication |
| K | Financial & insurance |
| L | Property |
| M | Professional, scientific and technical |
| N | Business, administration & support services |
| O | Public administration & defence |
| P | Education |
| Q | Health |
| R, S, T and U | Arts, entertainment & other services |

TABLE A5: Industrial Categories Used and Constituent Broad Industrial Group Sectors

| Industrial categories | Broad industrial groups |
|---|---|
| Knowledge intensive | Information & communication, Financial & insurance, Financial & insurance, Professional, scientific and technical and Education (J, K, M, N and P) |
| Manufacturing & construction | Manufacturing & Construction (C and F) |
| Arts, entertainment, recreation & other services | Arts, entertainment & other services (R, S, T and U) |
| Wholesale, transport & storage | Wholesale and Transport & storage (including postal) (G and H) |
| Public sector & health | Public administration & defence and Health (O and Q) |
| Retail & motor trades | Retail and Motor trades (G) |
| Accommodation, food services & property | Accommodation & food services and Property (I and L) |

| | Primary industries & utilities | Agriculture, forestry & fishing and Mining, quarrying & utilities (A, B, D and E) |